# Goos-Hänchen effect for Brillouin light scattering by acoustic phonons

Yuliya Dadoenkova,[1,2,*] Nataliya Dadoenkova,[1,2] Maciej Krawczyk,[3] Igor Lyubchanskii[2,4]

[1]*Ulyanovsk State University, Ulyanovsk 432000, Russia*
[2]*Donetsk Physical and Technical Institute of the NAS of Ukraine, Ukraine*
[3]*Faculty of Physics, Adam Mickiewicz University, Poznań, Poland*
[4]*V.N. Karazin Kharkiv National University, Kharkiv, Ukraine*
*\*Corresponding author: yulidad@gmail.com*

The lateral shift of an optical beam undergoing a Brillouin light scattering on acoustic wave in the total internal reflection geometry is studied theoretically. It is shown that the lateral shift depends on polarization (longitudinal or transversal) of the acoustic wave, as well as on the type of the scattering process: direct one, when the scattered wave undergoes a lateral shift at reflection from the interface, and cascading one, when fundamental-frequency light beam is laterally shifted at reflection and then is scattered on the acoustic wave. © 2015 Optical Society of America

http://dx.doi.org/10.1364/OL.99.099999

Effect of the lateral beam shift at the reflection of light from the interface between two media known as the Goos-Hänchen (GH) effect was observed for the first time in a glass [1] when the incidence angle of the light beam was close to the total internal reflection (TIR) angle. Until now this phenomenon is comprehensively studied in different materials (dielectrics, metals, magnetics, superconductors) [2-9] and complex structures (photonic crystals and multilayers) [10–12]. In all these papers and big number of other publications including several review articles [13-15], the GH effect has been studied in stationary cases when the incident and reflected beams were at the same frequencies.

In optically nonlinear media, the GH effect has been studied for materials with quadratic optical nonlinearity for three electromagnetic waves interaction, i.e. for the second harmonic generation [16, 17]. It should expect that it would be possible to observe the GH shift in the case of three-wave interaction when one of these waves is acoustic wave (AW). This type of coupling between the electromagnetic waves (EMWs) and AWs, which can be described via photoelastic interaction [18], leads to the well-known phenomenon of inelastic light scattering by sound (or acoustical phonons), i.e., Brillouin light scattering (BLS) [19, 20]. In this case, reflected wave at the frequency $\omega\pm\Omega$ will be also characterized by a lateral shift, where $\omega$ and $\Omega$ are the angular frequencies of EMW and AW, respectively. The influence of a static misfit strain at the film/substrate interface on the beam shifts due to photoelastic interaction has been done by us in [8, 21]. Generalization of this approach to study of the GH effect at BLS can be done by replacing the static (time-independent) strain contribution into permittivity tensor with the dynamic (time-dependent) strain produced by AWs.

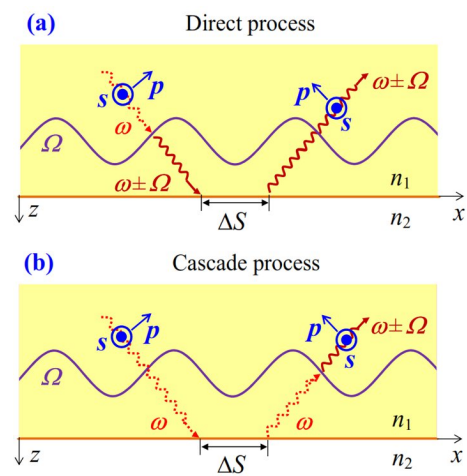

Fig. 1. Schematics of the direct (a) and cascade (b) BLS processes. Here **s** and **p** denote s- and p-polarized light, $\omega$ and $\Omega$ are fundamental light frequency and frequency of the acoustic wave, and $\Delta S$ is the GH shift.

In this Letter, in the framework of the phenomenological description we study the GH effect at BLS by AWs. To our best knowledge, this phenomenon has never been studied before. We investigate two possible ways (direct and cascade) to observe the GH effect for both longitudinal and transversal AWs. In the direct case, the incident light wave with the frequency $\omega$ interacts with the AW with the frequency $\Omega$, and the reflected EMW with the frequency $\omega \pm \Omega$ will undergo a lateral GH shift. In the cascading case, the incident light beam with the frequency $\omega$ is reflected from the interface, where the GH effect takes place, and after that the reflected EMW interacts with AW and the registered spatially-shifted radiation will be at the frequency $\omega \pm \Omega$, as in the direct process.

Let us consider thick film of cubic material located as shown at Fig. 1. Propagation of EMWs with the angular frequency $\omega$ is described by the solution of the wave equation [22]

$$\nabla \times \nabla \times \mathbf{E}(\mathbf{r},t) + \frac{n_1^2}{c} \frac{\partial^2 \mathbf{E}(\mathbf{r},t)}{\partial t^2} = -\mu_0 \frac{\partial^2 \mathbf{P}(\mathbf{r},t)}{\partial t^2} \quad (1)$$

where E(r,t) is the electric field of optical wave, $n_1$ is the refractive index, c is the velocity of light in vacuum, $\mu_0$ is the vacuum permeability, and P(r,t) is polarization vector.

An AW propagating in crystalline medium modulates its permittivity tensor via photoelastic interaction [22]. In this case, additional contribution $\Delta P$ to polarization can be presented in the form

$$\Delta P_i(\mathbf{r},t) = p_{ijkl} u_{kl}(\mathbf{r},t) E_j(\mathbf{r},t) \quad (2)$$

where $p_{ijkl}$ is the photoelastic tensor, and $u_{kl}$ is the strain tensor

$$u_{kl}(\mathbf{r},t) = \frac{1}{2}\left(\frac{\partial u_k(\mathbf{r},t)}{\partial r_l} + \frac{\partial u_l(\mathbf{r},t)}{\partial r_k}\right), \quad (r_k, r_l) \in \{x, y, z\} \quad (3)$$

Here u(r, t) is the displacement vector of travelling longitudinal (L) and transversal (T) AWs propagating along the x-axis with the angular frequencies $\Omega_{L,T}$ and wavenumbers $q_{L,T}$

$$\mathbf{u}_L(\mathbf{r},t) = \mathbf{A}_L e^{i(q_L x - \Omega_L t)}, \quad \mathbf{u}_T(\mathbf{r},t) = \mathbf{A}_T e^{i(q_T x - \Omega_T t)}, \quad (4)$$

where $\mathbf{A}_L = [0, 0, A_z]$ and $\mathbf{A}_T = [A_x, A_y, 0]$ are the magnitudes of these AWs.

The amplitude of the EMW modulated by the AW due photoelastic interaction can be found as a solution of wave equation with the polarization at combined frequencies:

$$\nabla \times \nabla \times \mathbf{E}(\mathbf{r},t) + \frac{n_1^2}{c} \frac{\partial^2 \mathbf{E}(\mathbf{r},t)}{\partial t^2} = -\mu_0 \frac{\partial^2 \Delta \mathbf{P}(\mathbf{r},t)}{\partial t^2} \quad (5)$$

Taking the Fourier transform of Eq. (5) with respect to time and using the slowly varying envelope approximation [23], where $\mathbf{k}_{\omega \pm \Omega} \cdot \nabla \mathbf{E}(\mathbf{r}, \omega \pm \Omega) \gg \Delta \mathbf{E}(\mathbf{r}, \omega \pm \Omega)$, we obtain the following reduced form or the wave equation for BLS process:

$$\mathbf{k}_{\omega \pm \Omega} \cdot \nabla \mathbf{E}(\mathbf{r}, \omega \pm \Omega) = -2i \frac{\omega^2}{c^2} \Delta \mathbf{P}(\mathbf{r}, \omega \pm \Omega) \exp(i \mathbf{k}_{\omega \pm \Omega} \mathbf{r}) \quad (6)$$

The electric field components $E_i(\omega \pm \Omega)$ of the scattered EMW can be obtained from the corresponding polarization terms through the following equation:

$$E_i(\omega \pm \Omega) = \frac{i\omega^2}{2c^2 k_{\omega \pm \Omega}} \left(\frac{1}{V} \int_V P_i(\omega \pm \Omega) e^{i\mathbf{Q} \cdot \mathbf{r}} \, dr\right). \quad (7)$$

where $\mathbf{Q} = \mathbf{k}_\omega + \mathbf{q} - \mathbf{k}_{\omega \pm \Omega}$ is the wave vectors mismatch.

Formation of the shifted scattered light beam of frequency $\omega \pm \Omega$ can contribute from two processes. In the direct BLS process, an EMW of fundamental frequency $\omega$ is scattered by an AW and the resulting wave of frequency $\omega \pm \Omega$ undergoes a lateral shift $\Delta S$ at reflection (see Fig. 1(a)). In the cascade process, first, a fundamental frequency beam is shifted at the reflection at the interface, and the reflected beam scatters by an AW (Fig. 1(b)).

Assuming an incident Gaussian beam of waist w$_0$, the spatial profile $E^{(sc)}(x)$ of the reflected (scattered) beam can be calculated as:

$$E^{(sc)}(x) = \frac{1}{2\pi} \int_{-\infty}^{\infty} \tilde{E}^{(i)}(K) R(K + k_c) \exp(iKx) dK,$$

$$\tilde{E}^{(i)}(K) = \int_{-\infty}^{\infty} E^{(i)}(x) \exp(-iKx) dx, \quad (8)$$

where $R(k_x)$ is reflection (scattering) function, $k_x$ is the component of the wavevector along the x-axis, $k_c$ is the central wavevector of the incident beam and $K = k_x - k_c$. In the direct BLS, $E^{(sc)}(x) = E^{(r)}_{\omega \pm \Omega}(x)$, and in the cascade BLS $E^{(sc)}(x) = E^{(r)}_\omega(x)$, where the superscript (r) refers to the wave reflected at the interface between media 1 and 2. Using Eq. (8), the intensity distribution of the scattered wave can be calculated. The position of its maximum will thus define the GH shift of the scattered light.

Numerical simulations are carried out for medium 1 and 2 to be ZnSe and vacuum, respectively. An incoming fundamental EMW of (unless other stated) the wavelength of He-Ne laser $\lambda_0$ = 633 nm which corresponds to the angular frequency in ZnSe $\omega = 2\pi c n_1 / \lambda_0 \approx 7.68 \cdot 10^{15}$ rad·Hz. The frequencies of the longitudinal and transverse AWs are $\Omega_L$ = 7.59 rad·kHz and $\Omega_T$ = 6.39 rad·kHz [24]. The refractive index of ZnSe $n_1(\omega) \approx n_1(\omega \pm \Omega_{L,T})$=2.578, and the angle of total internal reflection then is $\theta_{TIR}$ = asin(1/$n_1$) $\approx$ 22.8°. The photoelastic tensor components of ZnSe at $\lambda_0$ are $p_{11}$ = 0.017, $p_{12}$ = 0.078, and $p_{44}$ = 0.06 [25]. We assume the waist of the incoming Gaussian beam is w$_0$ = 15·$\lambda_1$ = 15·$\lambda_1$/$n_1$.

The symmetry of ZnSe allows to obtain S- (or P-) polarized scattered light from s- (or p-) polarized incoming light, i.e. s-S (or p-P) scattering, in the direct process whatever the type (longitudinal or transverse) of the AW. This is also the case for the cascade process for the longitudinal AW. For the cascade process in the case of the transverse AW, cross-polarization scatterings s-P and p-S are allowed, as well as p-P scattering, whereas s-S process is forbidden. It should be noted that the direct and cascade scattering by the longitudinal AW in the p-P and s-S processes takes place due to $p_{11}$ and $p_{12}$ photoelastic tensor components, respectively, whereas direct and cascade scattering by the transverse AW takes place due to $p_{44}$ for all polarization processes.

The intensities of the scattered beams normalized on the incoming beam intensity are shown in Figs. 2 and 3 for the direct and cascade process, respectively. The white lines show the position of the intensity maximum this defining the GH shift.

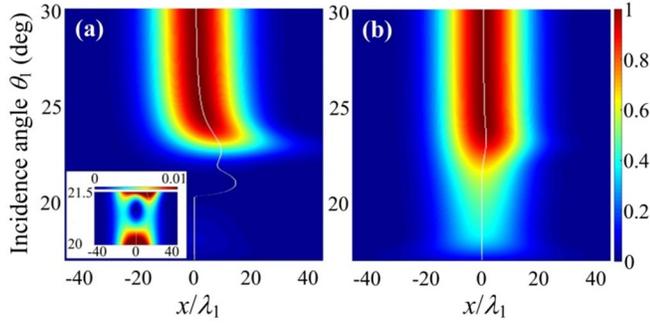

Fig. 2. Normalized intensity distribution of the scattered beam with the reduced spatial coordinate $x/\lambda_1$ and incidence angle $\theta_1$ in the direct BLS process with longitudinal and transverse AWs: (a) p-P scattering; (b) s-S scattering. The inset on (a) is a zoom of the intensity distribution around Brewster angle dip. The white line shows the position of the intensity maximum which defines the GH shift.

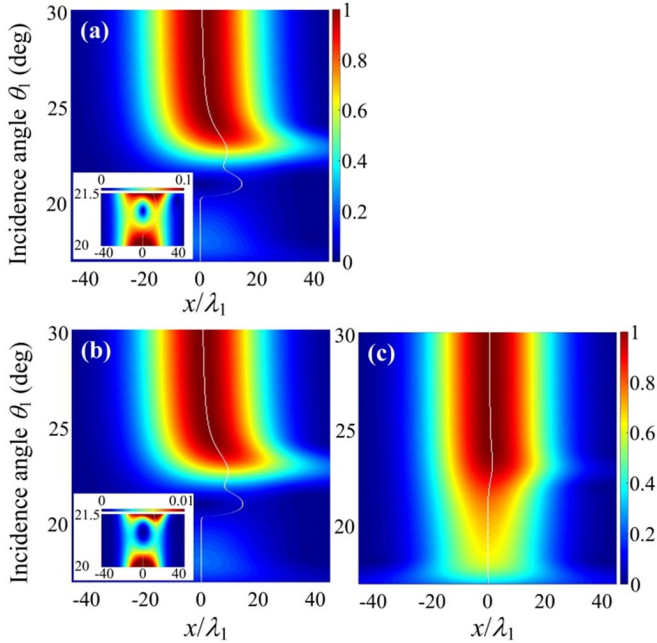

Fig. 3. Normalized intensity distribution of the scattered beam with the reduced spatial coordinate $x/\lambda_1$ and incidence angle $\theta_1$ in the cascading BLS process with transverse AW: (a) p-P scattering; (b) p-S scattering; (c) s-P scattering. The insets on (a) and (b) are zooms of the intensity distribution around Brewster angle dip. The white line shows the position of the intensity maximum which defines the GH shift.

The GH shift in the p-P scattering process demonstrates two maxima: first maximum around the Brewster angle $\theta_B = \mathrm{atan}(1/n_1) \approx 21.2°$, and the second one around $\theta_{TIR}$. The former case requires more detailed study, as the intensity of the scattered beam is minimal, as one can see from Fig. 2(a). However, the definition of the GH shift in the vicinity of the Brewster angle obscures, as the scattered beam undergoes strong reshaping and splits into two-peak pulse (see the inset on Fig. 2(a)). Similar reshaping of the scattered light takes place in the cascade BLS with the transverse AW in the p-S and p-P processes. In the latter polarization combination, the intensity of light is one order of magnitude larger than in the former one (compare the insets on Figs. 3(a) and 3(b)).

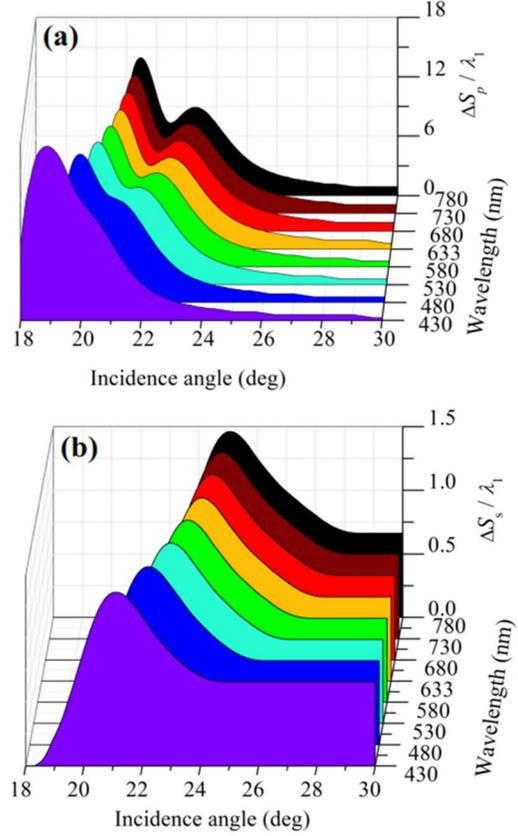

Fig. 4. Normalized GH shift $\Delta S/\lambda_1$ of the scattered light beams as functions of the incidence angles $\theta_1$ for different wavelength $\lambda_0$ for (a) p- and (b) s-polarized incident beam.

Refractive index of ZnSe is characterized by significant frequency dispersion in the visible regime. Figures 4(a) and 4(b) show angular dependencies of the GHS in the p-P and s-S processes for different wavelength $\lambda_0$. In that range, $n_1$ decreases with $\lambda_0$ increase, which results in shift of both $\theta_{TIR}$ and $\theta_B$ towards higher incidence angles, leading to the corresponding drift if the GHS maxima. For both p-P and s-S processes, maxima of the GHS decrease as $\lambda_0$ increases from the blue to the red light wavelengths. In the meantime, the GHS peak $\Delta S_p/\lambda_1$ at $\theta_{TIR}$ obscures for the blue light as it merges with the GH shift peak at $\theta_B$ (Fig. 4(a)). In the vicinity of $\theta_{TIR}$ the amplitude of $\Delta S_p/\lambda_1$ is about one order of magnitude larger than that of $\Delta S_s/\lambda_1$.

In conclusion, we have shown theoretically and numerically that at the Brillouin light scattering it is possible to observe an appearance of the lateral shift of reflected light at the frequencies shifted with the frequencies of acoustic waves. We showed that the lateral shift is sensitive to polarization state and incidence angle of the light beam, as well as to the acoustic wave polarization but in the cascade process only. One of possible methods to observe these shifts is using a micro-BLS spectroscopy [26] which allows registering small shifts of reflected light from the elementary excitations.

Funding. EU's Horizon 2020 MCSA-RISE (644348 (MagIC)); The Ministry of Education and Science of the Russian Federation (3.7614.2017/9.10, 14.Z50.31.0015)

Acknowledgment. We thank Y. S. Kivshar for fruitful discussion of this paper and useful comments.